\documentclass[a4paper,UKenglish,cleveref, autoref, thm-restate]{lipics-v2021}

\usepackage{listings}
\lstdefinelanguage{rtlola}{
  morekeywords={input,output,trigger,Float64,Bool},
  sensitive=true,
  morecomment=[l]{//},
  morestring=[b]"
}
\usepackage{tikz}
\usetikzlibrary{arrows.meta,positioning,calc}

\title{Experiential Learning of Runtime Monitoring Using Pachinko} 

\author{Miles Scharff}{Barnard College, Columbia University, New York City}{mscharff@barnard.edu}{https://orcid.org/0009-0000-5104-9179}{[NSF\#2525332]}

\author{Maria Chemodanova}{Barnard College, Columbia University, New York City}{mc5522@barnard.edu}{}{}

\author{Mark {Santolucito}}{Barnard College, Columbia University, New York City}{msantolu@barnard.edu}{https://orcid.org/0000-0002-1825-0097}
{[NSF\#2525332]}

\authorrunning{Scharff et al.} 

\Copyright{Miles Scharff, Maria Chemodanova, Mark Santolucito} 



\relatedversion{} 

\supplement{https://github.com/Barnard-PL-Labs/Runtime-Monitoring-Pachinko}

\nolinenumbers 

\EventNoEds{0}
\EventLongTitle{TEAL 2026: Tools For Educational Activities in Logic}
\EventShortTitle{TEAL 2026}
\EventAcronym{TEAL}
\EventYear{}
\EventDate{}
\EventLocation{}
\EventLogo{}
\SeriesVolume{}
\ArticleNo{}

\begin{document}

\maketitle

\begin{abstract}
We present documentation of a classroom assignment that teaches runtime monitoring through a creative embedded systems build: an interactive Pachinko game. The assignment centers on a dual-core ESP32 workflow in which students write RTLola specifications for monitors, compile these monitors to C, and deploy them alongside sensor and actuator control logic. Pachinko game events are logged in real time and used to trigger sound, animation, and motor behavior according to formal temporal logic specifications.

This work showcases how formal methods can be taught in a hands-on, project-based setting for learners in a creative and classroom-scale setting. We also discuss portability: the assignment template, hardware stack, code base, and assessment approach are designed and documented to be replicated in other embedded systems, creative computing, or makerspace-style courses. This assignment was given to the students of Creative Embedded Systems (COMS3930) at Barnard College.

\end{abstract}

\section{Introduction}
\label{sec:typesetting-summary}

Runtime monitoring is a formal verification technique used for real time critical cyber-physical systems such as drones or aircraft. It is often implemented as a separate module, the “monitor”, which receives data streams from other parts of the system and runs these streams through the monitor to guarantee that the data adheres to some formal correctness specification. 
The goal is to have a safety net so that if some subsystem fails, the runtime monitor is able to be a last resort system to catch such errors and raise an alarm to an appropriate handler. Runtime monitoring is especially useful in large, complex systems, such as flight systems or other systems considered mission-critical, that are worked on by many people or teams. As described in a NASA report on their runtime monitoring tool Copilot, they estimate the functional correctness of 10,000 lines of code would take a team of 20 engineers years to achieve~\cite{pike2013copilot},
hence runtime monitoring is a necessary solution for mission-critical systems. This context of large teams working on complex mission-critical systems is often what makes runtime monitoring so difficult to motivate in an educational setting.
The reality of classroom appropriate project scoping means the motivation for runtime monitoring is difficult to replicate.

\begin{figure}
    \centering
    \includegraphics[width=0.9\linewidth]{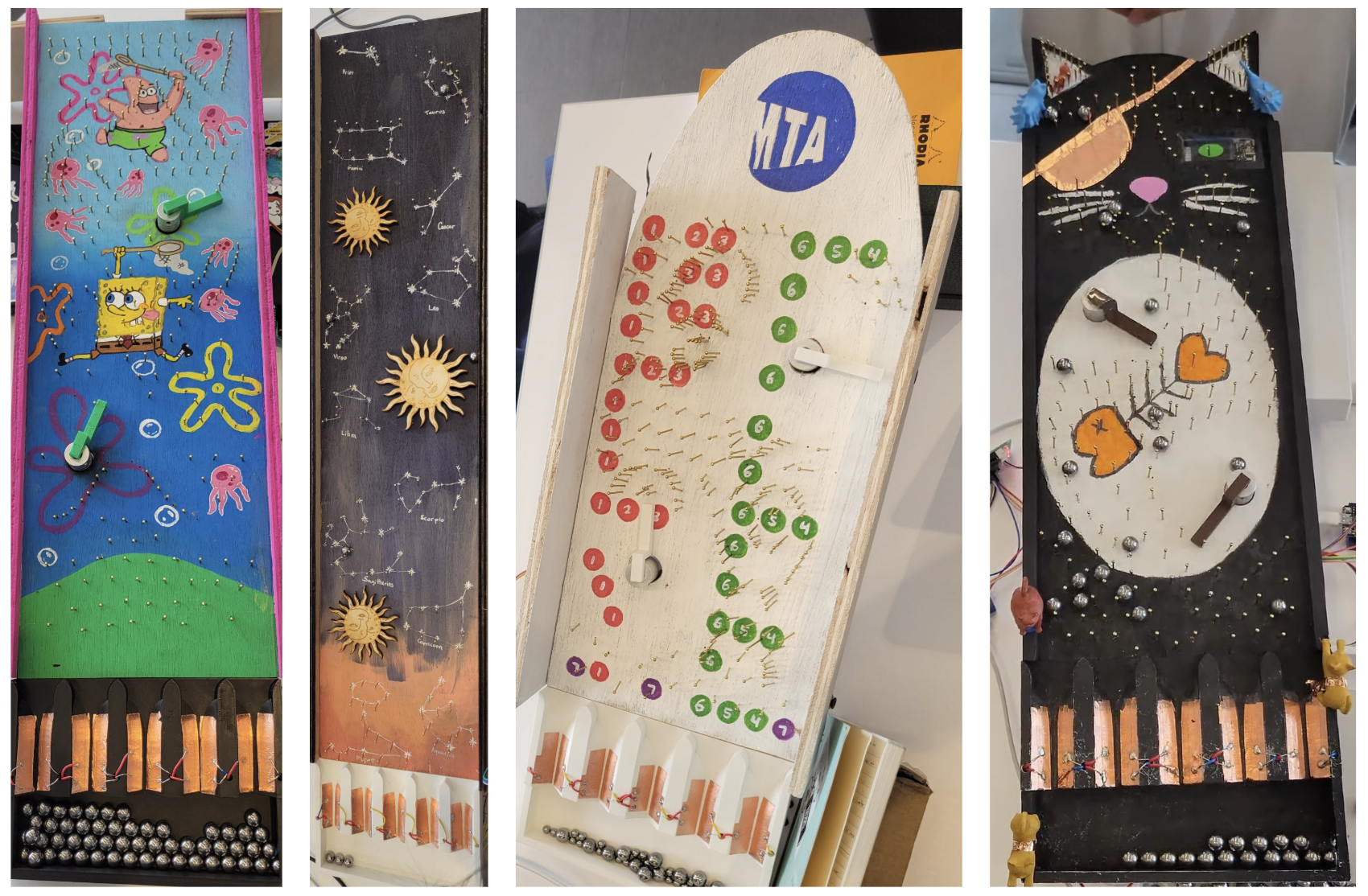}
    \caption{Student pachinko board projects from Spring 2026 in COMS3930 at Barnard College, Columbia University.}
    \label{fig:placeholder}
\end{figure}

We developed the Pachinko assignment to overcome these educational barriers and create an engaging and relevant introduction to runtime monitoring. Pachinko is a Japanese game involving steel balls falling down on to nails that determine their path. The goal of the player is to try and get as many balls into specific locations at the bottom of the board as possible. The assignment centers around an ESP32\footnote{The ESP32, developed by Espressif (\url{https://www.espressif.com/en/products/socs/esp32}), is an Arduino-like device that is often used in introductory embedded systems courses.} at the bottom of the Pachinko board that counts how many times a ball will roll into various positions. This ESP32 is running a C-compiled version of RTLola~\cite{d2005lola,faymonville2016stream,baumeister2025streamfairness}, a runtime monitoring tool, and pipes the behavior of these balls through formal specification. When a runtime monitor condition is triggered, the ESP32 sends messages to other the ESP32's on the board that control stepper motors to affect ball behavior. These messages are sent wirelessly through the ESP-NOW protocol \cite{espressif_espnow_espidf}.

In order for the students to complete the assignment successfully, they must write a formal specification in RTLola that triggers behaviors in their Pachinko game that satisfy their chosen artistic themes. 
In this context, system complexity is validated through artistic drive and simulated through the inherent noisy unpredictability of ball behavior in Pachinko. 
There is also a need to react to the ball distributions in real time. Hence, runtime monitoring and formal specifications feel well-motivated to use in this context.

Our goal for this first deployment of this project was to see how well students were able to write RTLola formal specifications to guide the desired behavior of their Pachinko game.

\section{Background and Related Work}

\subsection{Creative Embedded Systems Course}

The assignment was developed in the context of a broader curricular effort to integrate formal methods into creative embedded-systems education at Barnard College and Columbia University. That effort targets students building interactive media systems (e.g., sound, light, and installation pieces) on microcontroller platforms, where behavior is reactive, concurrent, and often distributed across multiple boards.

A key motivation is that creative-technology students increasingly use AI-assisted programming to build systems whose complexity can outpace their ability to reason about correctness. The curriculum therefore introduces lightweight formal methods directly in the toolchains students already use, including runtime verification \cite{leucker2011teaching,todorova2012runtime} with RTLola.

This pedagogical framing treats formal specification as practical design documentation for interactive behavior rather than an abstract prerequisite.
Prior course experience in creative embedded systems also provides natural settings for properties involving message passing, timing, and sensor-driven state changes, making the course a strong environment for introducing verification concepts to new communities.

\section{Pachinko Assignment}



Students are tasked with designing a Pachinko game that responds to steel-ball events by moving motors, producing sounds, or displaying animations.

The heart of the game is the ball logger, a 5 channel structure that allows balls to be detected if they pass through any of the individual channels. Each of the channels is lined with copper tape with a break in the middle. When a ball passes through a channel, it bridges the gap in the copper. One side of each channel is attached to the 3.3V pin on an ESP32 and the other side is attached to an analog input GPIO pin with a 10kOhm pull down resistor. The ESP32 is able to read whenever a ball bridges the two sides of a channel.

This ESP32 runs a C-compiled version of RTLola in a dual-core configuration using PlatformIO. Core 0 handles sensing and communication; core 1 runs the RTLola monitor. Core 0 sends sensor events to core 1, and core 1 returns flags when formal specifications are violated or fulfilled. Specifications are written in a \texttt{.lola} file, and a build script compiles each specification to a header included by the firmware. As a result, monitor updates are integrated into the normal embedded build process. 
Students are given a template for this system, so they can focus mainly on the specification itself.

\begin{lstlisting}[language=rtlola,caption={Example RTLola specification for repeated channel events.},label={lst:rtlola-pachinko}]
input sensor_1: Float64

output ball_1: Bool := sensor_1 > 0.5

output prev_ball_1: Bool := sensor_1.offset(by: -1).defaults(to: 0.0) > 0.5

trigger ball_1 && prev_ball_1 "Channel 1 repeated"
\end{lstlisting}

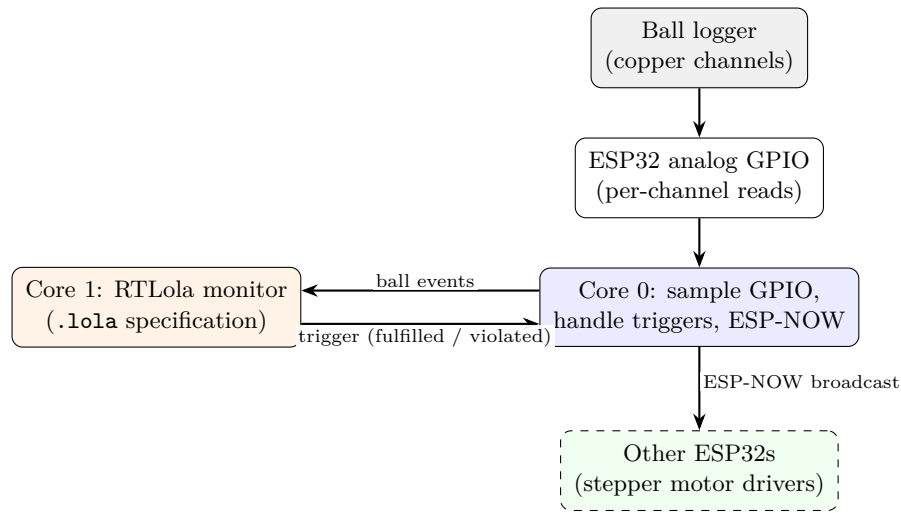
\begin{figure}[t]
  \centering
  \begin{tikzpicture}[
      node distance=0.65cm,
      box/.style={rectangle, draw, rounded corners, align=center, inner sep=5pt, font=\small},
      hw/.style={box, fill=black!6},
      c0/.style={box, fill=blue!8, minimum width=3.6cm},
      c1/.style={box, fill=orange!10, minimum width=3.6cm},
      ext/.style={box, dashed, fill=green!6},
      arr/.style={-{Stealth[length=2.2mm]}, thick},
      lbl/.style={font=\scriptsize, inner sep=1pt, fill=white, fill opacity=0.92, text opacity=1},
    ]
    \node[hw] (logger) {Ball logger\\(copper channels)};
    \node[box, below=of logger] (gpio) {ESP32 analog GPIO\\(per-channel reads)};
    \node[c0, below=of gpio] (c0) {Core 0: sample GPIO,\\handle triggers, ESP-NOW};
    \node[c1, left=3.15cm of c0] (c1) {Core 1: RTLola monitor\\(\texttt{.lola} specification)};
    \node[ext, below=1.1cm of c0] (remote) {Other ESP32s\\(stepper motor drivers)};
    \draw[arr] (logger) -- (gpio);
    \draw[arr] (gpio) -- (c0);
    \draw[arr] ($(c0.west)+(0,0.22)$) -- node[lbl, above, pos=0.48] {ball events} ($(c1.east)+(0,0.22)$);
    \draw[arr] ($(c1.east)+(0,-0.22)$) -- node[lbl, below, pos=0.52] {trigger (fulfilled / violated)} ($(c0.west)+(0,-0.22)$);
    \draw[arr] (c0) -- node[lbl, right, pos=0.42] {ESP-NOW broadcast} (remote);
  \end{tikzpicture}
  \caption{Signal flow on the central (monitor) ESP32: the ball logger is read on core~0, which exchanges events with the RTLola monitor on core~1; triggers return to core~0, which broadcasts actuation commands to distributed motor boards.}
  \label{fig:central-esp32-flow}
\end{figure}

In this example, \texttt{sensor\_1} is sampled from the logger hardware, \texttt{ball\_1} indicates current-channel occupancy, and \texttt{prev\_ball\_1} captures the prior sample. The trigger fires when occupancy is detected in consecutive samples, which we use as a simple repeated-event pattern before forwarding the corresponding action to the controller logic.

The students work in groups of 3-5 and are given a blank wooden board, stepper motors and shaft couplers, 500 brass nails, 200 stainless steel balls, copper tape, the 3D model for a ball logger, the 3D model for a motor arm, mounting screws for the ball logger and the motor arm, and a template PlatformIO project file with the RTLola dual core architecture set up. The template project also included some example formal specifications to get the students started. The total cost of materials for each board is approximately \$75 USD.

Before getting started, the students were asked to submit a work plan detailing the division of tasks within the group. The main tasks being writing of formal specification and trigger messages on the ``monitor'' ESP32, soldering of leads and resistors to the ball logger, testing stepper motor movements, planning nail locations, testing ESP-NOW communication between the boards, adding visual artistic elements, and assembly. 

\section{Reflections and Portability}

Students were able to write formal specifications to control the behavior of their boards with surprising ease, given the template PlatformIO project. The brunt of the issues and debugging students ran into was actually on the hardware side. Connecting a distributed system of microcontrollers together with sensors and motors provided some wiring challenges. Since runtime monitoring exists in complex embedded systems, this ended up being very relevant to real world usage. 

The main limitation of our method to run RTLola on an ESP32 is that RTLola's C compiler does not support temporal operators (e.g. \texttt{until} and \texttt{aggregate}). We could have opted to create our own runtime monitoring tool specifically for this assignment, but we felt that using an industry standard tool had more educational value. The students did have a desire to create temporal aggregates of ball counts, which was not supported by the current RTLola ESP32 framework. They were, however, able to achieve the same functionality by aggreating in C code on core 0, and sending the aggreate sums to the monitor. While this isn't ideal for teaching aggregation in a formal methods setting, the students were still able to think in formal specification techniques to achieve the desired behavior of their boards. We are working with RTLola to have the C-compiler receive temporal logic support in the future. 

Given the readily available and low cost physical materials for this project, as well as the ease of setup of the PlatformIO template project and pre-compiled RTLola C-compiler, we feel this assignment is not only portable but customizable. At it's core, the assignment is a distributed system of devices and motorized actuators controlled by a central monitor situated on a physical board. The number of distributed devices and motorized actuators and the shape of the physical board are easily variable, and hence so is the context that this assignment can be presented in a classroom. ESP-NOW supports up to 20 devices in unencrypted communication, meaning that this assignment could take the form of a full class collaborative assignment in some cases. Other instantiations of this assignment could include a wide range of other microcontroller sensors and actuators distributed across the board.

\bibliography{refs}

\end{document}